\documentclass[english]{article}
\usepackage[T1]{fontenc}
\usepackage[latin9]{inputenc}
\usepackage{mathrsfs}
\usepackage{amsmath}
\usepackage{amssymb}
\usepackage{esint}

\makeatletter

\newcommand{\noun}[1]{\textsc{#1}}

\makeatother

\usepackage{babel}
\begin{document}
\title{On the physical origin of the quantum operator formalism}
\author{A. M. Cetto, L. de la Peña and A. Valdés-Hernández}

\maketitle
Instituto de Física, Universidad Nacional Autónoma de México, \\04510
Mexico City
\begin{abstract}
We offer a clear physical explanation for the emergence of the quantum
operator formalism, by revisiting the role of the vacuum field in
quantum mechanics. The vacuum or random zero-point radiation field
has been shown previously---using the tools of stochastic electrodynamics---to
be central in allowing a particle subject to a conservative binding
force to reach a stationary state of motion. Here we focus on the
stationary states, and consider the role of the vacuum as a \emph{driving
force}. We observe that the particle responds resonantly to certain
modes of the field. A proper Hamiltonian analysis of this response
allows us to unequivocally trace the origin of the basic quantum commutator,
$\left[x,p\right]=i\hbar$, by establishing a one-to-one correspondence
between the response coefficients of $x$ and \emph{p} and the respective
matrix elements. The (random) driving field variables disappear thus
from the description, but their Hamiltonian properties become embodied
in the operator formalism. The Heisenberg equations establish the
dynamical relationship between the response functions. 
\end{abstract}

\section{Introduction}
\begin{flushright}
\textquotedbl Everything is still vague and unclear to me, but it
seems as if the electrons will no more move on orbits.\textquotedbl\cite{WH25}
\par\end{flushright}

Quantum mechanics with all its merits and successes, continues to
pose deep mysteries regarding the physics underlying the description.
A case in point is the operator formalism, the physical meaning of
which does not seem to have been essentially clarified since Heisenberg's
famous letter to Pauli dated June 1925, despite the impressive developments
of the theory. Given such lack of clarity, any serious effort to throw
light on this long standing matter warrants close attention. The present
paper is intended as a contribution to this effort. 

For this purpose we analyse the problem within the framework of nonrelativistic
quantum mechanics, from the perspective provided by stochastic electrodynamics,
\noun{sed} (see \cite{ClaDin80}-\cite{Boy rev} for reviews on \noun{sed}).
Developed on the premise that (otherwise classical) mechanical systems
are embedded in the vacuum or zero-point radiation field (\noun{zpf}),
taken as a \emph{real} field with an energy $\frac{1}{2}\hbar\omega$
per normal mode, \noun{sed} has produced a significant series of results
in accordance with quantum mechanics, suggesting that the action of
the field on matter is responsible for the emergence of quantization. 

In this work we revisit the effect of the \noun{zpf} on the particle
dynamics, with the aim to throw light on the physical origin and meaning
of the quantum operator formalism. Specifically, we focus our attention
on the \emph{driving} effect of the field on the particle once it
has reached a stationary state. In order to pave the way for the solution
of the general dynamical problem, we analyse first the essence of
the solution of the harmonic oscillator, the stationary part of which
is characterized by its resonant response to the \noun{zpf}. In the
general case of a nonlinear binding force, the system in a given stationary
state \emph{n} responds resonantly to a set of field modes of different
frequencies $\omega_{kn}$. By applying the appropriate Hamiltonian
treatment, on the basis that the field has taken control of the response,
we obtain the basic quantum commutator $\left[x,p\right]$ as a bilinear
expression involving the corresponding response functions, which are
identified with the matrix elements of the quantum operators. The
Heisenberg equations of motion for the operators follow as a natural
consequence. 

It should be noted that, although our point of departure is the equation
of motion for a charged particle in the presence of the \noun{zpf},
which is the usual starting point in \noun{sed}, the paper is self-contained
in the sense that to follow the arguments, the reader is not required
to be familiar with the details of \noun{sed} or with the process
leading to the quantum results obtained so far.

\section{Statement of the problem}

We recall that the usual starting point in \noun{sed} is the nonrelativistic
equation of motion for a particle of mass $m$ and charge $e$ (typically
an electron) subject to a conservative force $f(x)$ and to the electric
component of the random zero-point radiation field in the long-wavelength
approximation, $E(t)$. With the particle's radiation reaction included,
the equation reads \cite{ClaDin80,Dice} (we shall work throughout
in one dimension, for simplicity)
\begin{equation}
m\ddot{x}-f(x)-m\tau\dddot{x}=eE(t),\label{2}
\end{equation}
with \emph{$p=m\dot{x}$ }and $\tau=2e^{2}/3mc^{3}$. The \noun{zpf}
is a random stationary field with an energy $\hbar\omega/2$ per normal
mode. Since the energy of every mode is fixed, the only random element
is the phase of the mode; it is because of this randomness that the
problem becomes stochastic. The Fourier transform of $E(t)$ can therefore
be written as 
\begin{equation}
\frac{1}{2\pi}\intop dtE(t)e^{i\omega t}=\widetilde{E}(\omega)a(\omega),\label{4}
\end{equation}
 with 
\begin{equation}
\left|\widetilde{E}(\omega)\right|=\frac{1}{2\pi}\sqrt{\frac{\hbar\left|\omega\right|}{2}},\ a(\omega)=e^{i\phi},\label{6}
\end{equation}
and $\phi$ a random phase. Note incidentally that $a,a^{*}$ represent
precisely the normal variables conventionally used for the description
of a classical radiation field in reciprocal space (see, e. g., \cite{Cohen}).
The expression $a(\omega)$ is used here to stress that to each frequency
$\omega$ there corresponds a different field mode with an independent
random phase, without this implying a functional dependence on $\omega$. 

Let us first consider the specific case of a harmonic oscillator of
natural frequency $\omega_{0}$, for which Eq. (\ref{2}) can be solved
exactly. As is well known the solution $x(t)$ consists of two parts:
the transient one, decaying exponentially at a rate given by $\gamma=\tau\omega_{0}^{2}$,
and the stationary one, driven by the field $E(t)$. Oscillators,
atoms and other systems under consideration have been connected to
the field for times much longer than $\gamma^{-1}$, so that the transient
has disappeared and we are left with the stationary part of $x(t)$,
which is 
\begin{equation}
x(t)=\frac{e}{m}\intop d\omega\frac{\widetilde{E}(\omega)}{\omega_{0}^{2}-\omega^{2}-i\gamma\omega}a(\omega)e^{-i\omega t}+\mathrm{c.c}.\label{48}
\end{equation}

Notice from this equation that the oscillator's motion is under the
control of the \noun{zpf}, to which it responds linearly. Given the
fact that the ratio
\begin{equation}
Q=\frac{\omega_{0}}{\gamma}=\frac{1}{\tau\omega_{0}}\label{50}
\end{equation}
is very large for a typical atomic oscillator ($Q\sim10^{16}$ for
optical frequencies), the resonant response to the field is extremely
peaked. It is important to note that this resonant response is a property
of the mechanical system, i. e. of the oscillator embedded in the
\noun{zpf}, which will also manifest itself when it is excited by
an \emph{applied} field having a mode of frequency close to $\omega_{0}$;
in such case the oscillator may acquire an additional energy through
the excitation, thereby changing its state of motion.

\section{The nonlinear force problem}

We now extend the discussion to the general case of a particle being
subject to a nonlinear conservative binding force \emph{$f(x)$} and
having reached a stationary state of motion.\footnote{In \noun{sed} (\cite{ClaDin80}, or \cite{Dice} Ch. 5, it is shown
that such state of motion is reached as a result of the combined effect
of the background field and radiation reaction, as in the case of
the harmonic oscillator shown above. } The specificities of the possible stationary states of motion are
determined of course by $f(x)$. What is relevant to the present discussion
is that once the particle is in a stationary state, \emph{the }\emph{\noun{zpf}}\emph{
takes over and drives the response of the system}. As a typification
of the adduced situation, from solid-state physics we know that materials
subject to applied radiation fields---which are normally of high
intensity compared with the \noun{zpf---}respond linearly to such
fields, to a very good approximation.\footnote{Quoting from Ref. \cite{Pava}: ``Typically, the external (field)
force used in experiments is small with respect to the internal ones
(in a crystal), so that the system is weakly perturbed. Thus, the
dominant term is the linear response function. If we are able to disentangle
it, the linear-response function returns us information on the ground
state and the excitation spectrum, their symmetry properties, the
strength of correlations.'' It may of course happen that the intensity
of the applied field is so high (as is the case with current high-intensity
laser pulses) that the response of the system to it becomes nonlinear.
This case falls beyond the scope of the present discussion.} We may therefore safely assume that the particle responds also linearly
to the zero-point component of the radiation field. 

The appropriate tools to study the response of the system to the influence
of a stationary driving force are provided by the linear-response
theory (\noun{lrt}). An analysis of the response to different frequency
components of this force is known to provide important information
about the properties of the system itself \cite{four}. Guided by
the \noun{lrt}, for our system in a stationary state we write the
change produced by the driving force $eE(t)$ on the dynamical variable
$x$ as
\begin{equation}
x(t)=\frac{e}{m}\intop_{-\infty}^{+\infty}dt'\chi(t-t')E(t').\label{A4}
\end{equation}
Given the stationarity of the problem it is convenient to express
(\ref{A4}) in terms of the Fourier component of the stimulus, 
\begin{equation}
\widetilde{x}(\omega)=\frac{e}{m}\widetilde{\chi}(\omega)\widetilde{E}(\omega)a(\omega).\label{A6}
\end{equation}
This is a central proposal, indicating that the response of the system
is local in the reciprocal space. The validity of this expression
is verified by direct application of the property of Maxwell\textquoteright s
equations coupled to matter by charges and currents, whose four-dimensional
Fourier transform is local in the reciprocal space $k$ \cite{Cohen};
here we are applying it to the fourth component of $k_{\mu}$, namely
the frequency. 

In the case of the harmonic oscillator, the response function $\widetilde{\chi}(\omega)$
is peaked around its natural frequency $\omega_{0}$, which is its
single resonance frequency, as indicated by Eq. (\ref{48}). In the
general case, given the nonlinearity of the force $f(x)$ one should
expect that the particle is able to respond to more than one frequency,\footnote{In fact, materials are known to resonate in general to a series of
frequencies, but the responses are usually analysed separately, for
simplicity. } whence we write the response function as $\widetilde{\chi}(\omega)=\sum_{k}\widetilde{\chi}_{k}(\omega)$,
where
\begin{equation}
\widetilde{\chi}_{k}(\omega)=\frac{1}{\omega_{k}^{2}-\omega^{2}-i\gamma_{k}\omega}\label{A12}
\end{equation}
and the set of frequencies $\left\{ \omega_{k}\right\} $ is determined
by the specific form of $f(x)$, as will become clear later. The (inverse)
Fourier transform of the response function is then
\begin{equation}
\chi(t)=\frac{1}{2\pi}\intop_{-\infty}^{+\infty}d\omega e^{-i\omega t}\sum_{k}\widetilde{\chi}_{k}(\omega),\label{51}
\end{equation}
with $\mathrm{Re}\widetilde{\chi}_{k}(\omega)$ representing the reactive
terms, and $\mathrm{Im}\widetilde{\chi}_{k}(\omega)$ the dissipative
or absorptive terms.\emph{ }

According to the Kramers-Kronig relation \cite{four},
\begin{equation}
\widetilde{\chi}_{k}(\omega)=\intop_{-\infty}^{+\infty}\frac{d\omega'}{\pi}\frac{\mathrm{Im}\widetilde{\chi}_{k}(\omega')}{\omega'-\omega_{k}-i\gamma_{k}\omega},\label{A14}
\end{equation}
the higher the reactive response to a given frequency $\omega_{k}$,
the more intense tends to be the participation of the system in absorption
or emission processes at that frequency. This means that the system
is able to absorb or emit energy when it is perturbed by sources that
contain frequencies close to $\omega_{k}$. The new state reached
by the system will of course depend on the specific absorption or
emission taking place, i. e., on the frequency $\omega_{k}$. Therefore,
the field mode denoted by $k$ connects the initial and final states.

Since the response functions are highly peaked, we may replace them
with their value at the resonance frequency and write instead of (\ref{A4})
\begin{equation}
x(t)=\frac{e}{m}\sum_{k}\chi_{k}E_{k}a_{k}e^{-i\omega_{k}t}+\mathrm{c.c}.,\label{A16}
\end{equation}
where $\chi_{k}$ is the coefficient of the response to the field
mode of frequency $\omega_{k}$, and 
\begin{equation}
E_{k}(\omega_{k})=\frac{1}{2\pi}\sqrt{\frac{\hbar\left|\omega_{k}\right|}{2}}a_{k},\label{A18}
\end{equation}
with $a_{k}=e^{i\phi_{k}}$.

\section{Accounting for more than one stationary state}

When the dynamics of the system admits more than one stationary state---which
is generally the case for quantum systems---the set of response coefficients
to different frequencies, $\left\{ \chi_{k}\right\} $, will depend
in principle on the state. To specify the state we add an index $n$
to $x(t)$ and to the respective response coefficients. Equation (\ref{A16})
becomes thus
\begin{equation}
x_{n}(t)=\frac{e}{m}\sum_{k}\chi_{nk}E_{nk}a_{nk}e^{-i\omega_{kn}t}+\mathrm{c.c}.,\label{56}
\end{equation}
where 
\begin{equation}
E_{nk}(\omega_{kn})=\frac{1}{2\pi}\sqrt{\frac{\hbar\left|\omega_{kn}\right|}{2}}a_{nk},\label{58}
\end{equation}
and 
\begin{equation}
a_{nk}=e^{i\phi_{nk}}.\label{59}
\end{equation}
In Eq. (\ref{56}) the reverse order of the subindices in $\omega$
has been chosen in order to adjust to the normal convention of quantum
mechanics. At this point it is convenient to introduce the response
coefficients $x_{nk}$, normalized according to Eq. (\ref{58}),
\begin{equation}
x_{nk}=\frac{e}{2\pi m}\sqrt{\frac{\hbar\left|\omega_{kn}\right|}{2}}\chi_{nk},\label{60}
\end{equation}
so that Eq. (\ref{56}) takes the form
\begin{equation}
x_{n}(t)=\sum_{k}x_{nk}a_{nk}e^{-i\omega_{kn}t}\mathrm{+c.c.,}\label{61}
\end{equation}
with $\left|a_{nk}\right|=1$, according to Eq. (\ref{59}). Since
both $x_{n}(t)$ and $E(t)$ are real quantities, the respective coefficients
satisfy
\begin{equation}
x_{nk}^{*}(\omega_{kn})=x_{nk}(-\omega_{kn}),\ a_{nk}^{*}(\omega_{kn})=a_{nk}(-\omega_{kn}).\label{64}
\end{equation}
To complete the picture, from Eq. (\ref{61}) and 
\begin{equation}
p_{n}=m\dot{x}_{n}=-im\sum_{k}\omega_{kn}x_{nk}a_{nk}e^{-i\omega_{kn}t}+\mathrm{c.c}.\label{66}
\end{equation}
 we have
\begin{equation}
p_{nk}=-im\omega_{kn}x_{nk};\ p_{nk}^{*}=im\omega_{kn}x_{nk}^{*}=p_{nk}(-\omega_{kn}).\label{68}
\end{equation}

Let us for a moment consider a system with just two stationary states.
According to our observation following Eq. (\ref{A14}), the field
mode of frequency $\omega_{kn}$ connects state $n$ with state $k$.
Therefore, if the system is instead in state $k$, the mode of frequency
$\omega_{nk}$ connects it with state $n$, which implies that the
two indices $n,k$ enter on an equal footing in the above formulas.
The response coefficient $x_{nk}$ (or $x_{kn}$) determines whether
a transition $n\rightarrow k$ (or $k\rightarrow n)$ can take place;
if the transition $n\rightarrow k$ involves an absorption of energy
$\mathcal{E}_{nk}$, the same amount of energy is emitted in the transition
$k\rightarrow n$, i.e., $\mathcal{E}_{kn}=-\mathcal{E}_{nk}$.

We note incidentally that the present derivation of Eqs. (\ref{61})-(\ref{68})
confirms the legitimacy of the expressions introduced in the framework
of linear stochastic electrodynamics\noun{, lsed (}see \cite{ldlp avh amc 09},
or \cite{TEQ} Ch. 5). There the results followed from the demand
of ergodicity, whereas here they result from locality in frequency
space, Eq. (\ref{A6}). They tell us that for a given (binding) force
$f(x)$, and in a given (stationary) state of motion $n$, the system
responds resonantly and piecemeal to a given set of frequencies $\left\{ \omega_{kn}\right\} $,
with response coefficients $\left\{ x_{nk}\right\} $. In other words,
the set $\left\{ x_{nk},\omega_{kn}\right\} $ characterizes the problem
and the state of the system. In order to determine their values we
need to derive the corresponding dynamical equations. 

To clarify a confusion expressed from time to time, we stress that
the force $f(x)$, which is nonlinear in general, is \emph{not} being
approximated by a linear force; it is clearly the \emph{response}
of the system to the driving field $E(t)$ that is linear.

\section{Algebraic structure of the solution\label{Alg}}

If we were dealing with a classical dynamical system, consisting simply
of a particle subject to the force $f(x)$, the algebraic structure
would be that of the phase-space variables $(x,p)$, which define
the Poisson bracket
\begin{equation}
\left\{ x(t),p(t)\right\} _{xp}=1.\label{72}
\end{equation}
In the present case, by contrast, we are dealing with the responses
$x_{n}(t)$, $p_{n}(t)$ of the system in a given state $n$ to a
set of relevant field modes $\left\{ nk\right\} $. As indicated by
Eq. (\ref{61}), these responses depend directly on the field coefficients
$a_{nk}$, which are related to the phase-space variables of the respective
field modes, namely on $\left\{ \mathrm{q}_{nk},\mathrm{p}_{nk}\right\} $,
as follows,
\begin{equation}
\mathrm{q}_{nk}=\sqrt{\frac{\hbar}{2\left|\omega_{kn}\right|}}(a_{nk}+a_{nk}^{*}),\ \mathrm{p}_{nk}=-i\sqrt{\frac{\hbar\left|\omega_{kn}\right|}{2}}(a_{nk}-a_{nk}^{*}).\label{106}
\end{equation}
The algebraic structure of the space of the particle's response variables
must reflect this dependence on the set $\left\{ nk\right\} $ of
canonical field variables. This means that the Poisson bracket of
$x_{n}(t)$ and $p_{n}(t)$ shall not be taken with respect to $\left\{ x_{i},p_{i}\right\} $
(or $\left\{ x,p\right\} $ in the one-dimensional case), but with
respect to $\left\{ \mathrm{q}_{nk},\mathrm{p}_{nk}\right\} $. We
therefore write, for \emph{any} stationary state $n$,
\begin{equation}
\left\{ x_{n}(t),p_{n}(t)\right\} _{\mathrm{qp}}=\sum_{k}\left(\frac{\partial x_{n}}{\partial\mathrm{q}_{nk}}\frac{\partial p_{n}}{\partial\mathrm{p}_{nk}}-\frac{\partial p_{n}}{\partial\mathrm{q}_{nk}}\frac{\partial x_{n}}{\partial\mathrm{p}_{nk}}\right)=1,\label{74}
\end{equation}
where the subindex $\mathrm{qp}$ denotes the relevant set of canonical
field variables. The transformation rules (\ref{106}) applied to
this Poisson bracket allow to express it in terms of the normal field
coefficients, 
\begin{equation}
\left\{ x_{n}(t),p_{n}(t)\right\} _{\mathrm{qp}}=-\frac{i}{\hbar}\sum_{k}\left(\frac{\partial x_{n}}{\partial a_{nk}}\frac{\partial p_{n}}{\partial a_{nk}^{*}}-\frac{\partial p_{n}}{\partial a_{nk}}\frac{\partial x_{n}}{\partial a_{nk}^{*}}\right)=1.\label{76}
\end{equation}
Therefore, this new bilinear form, which we denote with a square bracket\footnote{This bilinear form was introduced already in \cite{PeCe86}, under
the name \emph{Poissonian}; see also \cite{Dice} Ch. 10).} 
\begin{equation}
\sum_{k}\left(\frac{\partial x_{n}}{\partial a_{nk}}\frac{\partial p_{n}}{\partial a_{nk}^{*}}-\frac{\partial p_{n}}{\partial a_{nk}}\frac{\partial x_{n}}{\partial a_{nk}^{*}}\right)\equiv\left[x_{n},p_{n}\right],\label{78}
\end{equation}
must, according to Eq. (\ref{76}), satisfy the equation
\begin{equation}
\left[x_{n},p_{n}\right]=i\hbar,\label{79}
\end{equation}
with $x_{n}$, $p_{n}$ given by (\ref{61}) and (\ref{66}). Since
the value of this Poisson bracket does not depend on the state $n$,
we may write in general
\begin{equation}
\left[x,p\right]=i\hbar\label{90}
\end{equation}
 for $x,p$ in any state.

This result is crucial: it indicates that \emph{the scale of the response
of the system to the field variables, expressed through the bilinear
form }(\ref{78})\emph{, is universally determined by Planck's constant,
that is, by the scale of the energy of the }\emph{\noun{zpf}}\emph{
modes}. Notice that the energy per normal mode $\frac{1}{2}\hbar\omega$,
expressed through the factor $\sqrt{\left|\omega_{nk}\right|/2\hbar}$
intervening in the transformation from canonical to normal variables---or,
equivalently, the spectral composition of the \noun{zpf} proportional
to $\omega^{3}$, which ensures its Lorentz invariance \cite{EH1910,TWM1963}---plays
a decisive role. Any other spectral form of the field would have led
to an explicit dependence on $\omega$ in (\ref{78}), which then
would not have a universal value.

To close this section we note that by using Eqs. (\ref{61})-(\ref{68}),
the binary operation on the variables $(x,p)$ denoted by the bracket
$\left[x,p\right]$ becomes expressed in terms of the response coefficients
$x_{nk}$ and the respective frequencies $\omega_{nk}$,
\begin{equation}
{\displaystyle \sum_{k}}\left(x_{nk}p_{nk}^{*}-p_{nk}x_{nk}^{*}\right)=2i{\displaystyle m\sum_{k}}\omega_{kn}\left|x_{nk}\right|^{2}=i\hbar,\label{92}
\end{equation}
whence

\begin{equation}
2m{\displaystyle \sum_{k}}\omega_{kn}\left|x_{nk}\right|^{2}=\hbar\label{94}
\end{equation}
 for any state $n$.

The process leading to (\ref{90}) reveals thus one of the most intricate
quantum enigmas, by exhibiting the Poisson bracket as the result of
the transition from the classical+\noun{zpf} dynamics to the quantum
dynamics, meaning the (Hamiltonian) dynamics describing the response
of the system to the driving field modes. Let us now build on this
result.

\section{Matrix mechanics; operators and the Hilbert space}

As observed in the paragraph following Eq. (\ref{68}), the system
is able to absorb or emit energy when it is perturbed by (field) sources
that contain the resonance frequencies $\omega_{nk}$. This means
that the response coefficients $x_{nk}$ connect a stationary state
$n$ with a stationary state $k$, of a higher or a lower energy depending
on the sign of $\mathrm{Im}\widetilde{\chi}_{nk}(\omega)$, given
by Eq. (\ref{A12}). Therefore, the sign of $\omega_{nk}$ reverses
with the interchange of $n$ and $k$, i.e., 
\begin{equation}
\omega_{nk}=-\omega_{kn}.\label{95}
\end{equation}
Further, from this equation and Eqs. (\ref{64}) we get
\begin{equation}
x_{nk}^{*}(\omega_{nk})=x_{kn}(\omega_{kn}),\ p_{nk}^{*}(\omega_{nk})=p_{kn}(\omega_{kn}),\ a_{nk}^{*}(\omega_{nk})=a_{kn}(\omega_{kn}).\label{96}
\end{equation}
Therefore Eq. (\ref{92}) can be written, as was promptly discovered
by Born, 
\begin{equation}
{\displaystyle \sum_{k}}\left(x_{nk}p_{kn}-p_{nk}x_{kn}\right)=i\hbar,\label{98}
\end{equation}
and its expression in the form of Eq. (\ref{94}) coincides then with
the Thomas-Reiche-Kuhn sum rule.

The coefficients $x_{nk}$ ($p_{nk}$) have thus become the elements
of a matrix $\mathbb{X}$ ($\mathbb{P}$) with as many rows and columns
as there are different states, and Eq. (\ref{98}) can be written
in terms of the matrix products, 
\begin{equation}
\left(\mathbb{XP}-\mathbb{PX}\right)_{nn}=i\hbar\mathbb{I}_{nn},\label{100}
\end{equation}
for any state $n$. Since the normal field variables connecting different
states $n,n'$ are independent, we have $\left(\partial a_{nk}/\partial a_{n'k}\right)=\delta_{nn'}$;
this allows us to write Eq. (\ref{100}) in the more general form
\begin{equation}
\left(\mathbb{XP}-\mathbb{PX}\right)_{nn'}=i\hbar\delta_{nn'},\label{102}
\end{equation}
which is just the matrix formula discovered by Heisenberg for the
quantum rule, in fact the basic quantum commutator.

Therefore, the \emph{basic quantum commutator is the matrix expression
of the Poisson bracket of the system's response variables with respect
to the normal field variables.} The matrix element $x_{nk}$ represents
the (resonant) response amplitude of the system in state $n$ to the
field mode $\left(nk\right)$.

Since the matrix elements connect different stationary states $n$
and $k$, the corresponding vectors are precisely the state vectors,
and the dimension of the Hilbert space is determined by the total
number of possible (stationary) states. Diagonal matrices represent
conserved quantities which do not connect different states; the state
vectors are therefore orthogonal. For the one-dimensional conservative
problem studied here, the only conserved quantity is the energy, represented
by the Hamiltonian operator 
\begin{equation}
\mathbb{H}=\frac{\mathbb{P}^{2}}{2m}+\mathbb{V}(\mathbb{X}).\label{110}
\end{equation}

With the help of the Leibniz rule and using Eqs. (\ref{102}) and
(\ref{110}), it is a simple exercise to derive the equations of evolution
for the matrices $\mathbb{X}$, $\mathbb{P}$. For the first one we
obtain, using Eq. (\ref{66}),
\begin{equation}
\frac{1}{i\hbar}\left[\mathbb{X},\mathbb{H}\right]=\frac{\mathbb{P}}{m}=\mathbb{\dot{X}}.\label{104}
\end{equation}
For the second one we get
\begin{equation}
\frac{1}{i\hbar}\left[\mathbb{P},\mathbb{H}\right]=\mathbb{F}=\mathbb{\dot{P}}.\label{112}
\end{equation}
In writing the second equality we have taken into account that in
the stationary states, the radiation reaction appearing in Eq. (\ref{2})
is balanced by the \noun{zpf}, as observed in connection with the
harmonic oscillator, which allows us to write $f_{n}(t)=m\ddot{x}_{n}(t)=\dot{p}_{n}(t)$.

To close the loop we use the fact that in its own representation $\mathbb{H}$
is diagonal, as stated previously, and write $H_{nk}=\mathscr{E}_{n}\delta_{nk}$;
whence from Eqs. (\ref{66}) and (\ref{104}) we obtain the well-known
Bohr formula for the energy involved in the transition between states
$k$ and $n$,
\begin{equation}
(\mathscr{E}_{k}-\mathscr{E}_{n})=\hbar\omega_{kn}.\label{114}
\end{equation}

\section{Corollary}

The matrix formulation of quantum mechanics constitutes a most compact
and elegant means of describing the (resonant) response properties
of a particle subject to a (binding) conservative force $f=-V'(x)$,
embedded in the \noun{zpf} and under its control. The random vacuum
field has introduced a fundamental change in the description of the
dynamical behavior of the system \cite{CePeVa2020}, reflected in
the transition from the classical to the quantum Poisson bracket.
Regardless of the form of the potential $V(x)$, the particle's response
to the field, expressed through $\left(x_{n}(t),p_{n}(t)\right)$,
is linear and becomes represented by a set of harmonic oscillators,
the amplitudes and frequencies of which depend on the external force
and on the state of the system. The resonance frequencies are precisely
the spectral frequencies, and the response amplitudes are the corresponding
matrix elements. It is clear that the pair $\left(x_{n}(t),p_{n}(t)\right)$
does not refer to the trajectory of the particle, whence there is
no (true) phase-space distribution associated with it. 

The migration of the description from the physical space to the abstract
Hilbert space offers a physical explanation for the latter, as Lorentz,
Schrödinger, and other pioneers of the quantum theory were eagerly
reclaiming at the time \cite{Bacci}. Heisenberg was of course right
in stating that his matrix mechanics does not describe the orbital
motion of electrons; but he did not have at hand the vacuum field
discovered by Planck fifteen years earlier, to develop his theory
on more physical grounds.

In a more elaborate version we shall discuss the physical implications
of the results presented here.

\end{document}